\documentclass[12pt]{amsart}

\numberwithin{equation}{section}
\newtheorem{definition}{Definition}[section]

\newtheorem{theorem}{Theorem}[section]
\newtheorem{proposition}{Proposition}[section]
\newtheorem{lemma}{Lemma}[section]

\theoremstyle{remark}
\newtheorem{remark}{Remark}[section]

\newcommand{\sub}[2]{\underset{{#1}}{\underbrace{{#2}} } }

\newcommand{\scal}[2]{\bigl\langle {#1} , {#2} \bigl\rangle}
\newcommand{\dx}{\,\text{d}x}
\newcommand{\dy}{\,\text{d}y}

\newcommand{\ds}{\,\text{d}s}
\newcommand{\ep}{\varepsilon}
\newcommand{\vare}{\varepsilon}
\newcommand{\wt}{\widetilde}
\newcommand{\wb}{\overline}
\newcommand{\al}{\alpha}
\newcommand{\la}{\lambda}
\renewcommand\phi{\varphi}
\newcommand\nlp[2]{\|{#2}\|_{L^{{#1}}}}
\newcommand{\paragraf}[1]{\vskip\baselineskip\indent{\bfseries {#1}.}}
\newcommand{\loc}{{\scriptsize \mbox{loc}}}
\DeclareMathOperator\dive{div}
\DeclareMathOperator\curl{curl}
\DeclareMathOperator\supp{supp}
\DeclareMathOperator\logt{\wt{log}}
\newcommand\omt{\,\wt\omega}

\newcommand\omty{\omt(t,y)}
\newcommand\omtxx{\,\omega(t,x)}
\newcommand\omtyy{\,\omega(t,y)}

\newcommand{\R}{\mathbb{R}}
\newcommand{\N}{\mathbb{N}}
\newcommand{\HH}{\mathbb{H}}
\newcommand\dert{\frac{\text d}{{\text d}t}}
\begin{document}
\begin{center}
\large\bf
LARGE TIME BEHAVIOR FOR VORTEX EVOLUTION IN THE HALF-PLANE
\end{center}
\vskip\baselineskip
\begin{center}
\sc 
D. Iftimie\\
M.C. Lopes Filho\footnote{Research supported in part by CNPq grant \#300.962/91-6}\\
H.J. Nussenzveig Lopes\footnote{Research supported in part by CNPq grant \#300.158/93-9}  
\end{center}
\vskip2\baselineskip
\begin{center}
\parbox{14cm}{\scriptsize
{\sc Abstract.} 
 In this article we study the long-time behavior of incompressible ideal flow in a half plane from the point of view of vortex scattering. Our main result is that certain asymptotic states for half-plane vortex dynamics decompose naturally into a nonlinear superposition of soliton-like states. Our approach is to combine techniques developed in the study of vortex confinement with weak convergence tools in order to study the asymptotic behavior of a self-similar rescaling of a solution of the incompressible 2D Euler equations on a half plane with compactly supported, nonnegative initial vorticity.
}\end{center}
\vskip1cm

\tableofcontents

\section{Introduction}

        Let $\omega = \omega(t,x)$ be the vorticity associated to a solution of the incompressible two-dimensional Euler equations on the upper half-plane with an initial vorticity $\omega_0$ which is bounded, compactly supported and nonnegative. We consider a rescaling $W = W(t,x) = t^2\omega(t,tx)$, whose time-asymptotic behavior encodes information on the scattering of $\omega$ into traveling wave solutions of the 2D Euler system on the half-plane. This choice of rescaling was also made in view of the fact that the horizontal velocity of the center of vorticity is bounded away from zero from below (see \cite{Iftimie999999}). The rescaling $W$ is weakly compact as a time-dependent family of measures. The main purpose of this work is to present a structure theorem, stating that if the rescaling $W$ is actually weakly convergent to a measure then this measure must be of the form $\sum m_i \delta(x_1-\alpha_i) \otimes \delta(x_2)$, with $m_i > 0$, $\alpha_i$ a discrete set of points on an interval of the form $[0,M]$ whose only possible accumulation point is $x_1=0$, and where $\delta$ denotes the one-dimensional Dirac measure centered at $0$.

        Let us begin with a precise formulation of vortex dynamics on the half-plane. Let $(x_1,x_2)$ be the coordinates of a point $x$ in the plane and denote the upper half-plane by ${\Bbb H} \equiv \{x_2>0\}$. The initial-boundary value problem for the incompressible 2D Euler equations in ${\Bbb H}$ is given by:
\begin{equation} \label{vorteq}
\left\{ \begin{array}{l}
\omega_t + u \cdot \nabla \omega = 0, \mbox{ in } (0,\infty)\times \Bbb H  \\
\mbox{div }u = 0, \mbox{ in } [0,\infty)\times \Bbb H  \\
\mbox{curl }u = \omega, \mbox{ in } [0,\infty)\times \Bbb H \\
u_2(t, x_1,0) = 0, \mbox{ on } [0,\infty)\times \R  \\
\omega(0,x_1,x_2) = \omega_0(x_1,x_2) \mbox{ at } \{t=0\}\times \Bbb H  , \end{array} \right. \end{equation} 
with $u=(u_1,u_2)$ the velocity and $\omega$ the vorticity of the flow.

For bounded, compactly supported initial vorticity, problem (\ref{vorteq}) is equivalent to the full plane problem with initial vorticity given by an odd extension of $\omega_0$ to $\{x_2<0\}$ (see \cite{LNX01} for details). Global well-posedness of the initial boundary value problem follows from this equivalence, using Yudovich's Theorem \cite{yudovich63}. For compactly supported initial vorticity in $(L^1+{\mathcal BM}_+)\cap H^{-1}_{\loc}$ we have global
existence of weak solutions by adapting Delort's Theorem to the half-plane case, see \cite{delort91,LNX01,schochet95,VW93}, but uniqueness is open.   

The present work is best understood within the context of research on
vorticity confinement. Let $\omega = \omega(t,x)$ be a (classical)
solution of the full plane 2D Euler equations such that $\omega(0,x)$ is
compactly supported. The problem of confinement of vorticity is to
understand the spreading of the support of $\omega(t,\cdot)$ for large
time. The main result in confinement of vorticity states that if the
initial vorticity is nonnegative, with support contained in the ball
$B(0;R_0)$, then for any $a > 1/4$  there exists $b>0$ such that the 
support of vorticity at time $t$ is contained in a ball of radius $R(t)=(bt +
R_0^{1/a})^a$. This result is due to C. Marchioro \cite{marchioro94} for
$a=\frac13$ and was improved to $a>\frac14$ in \cite{ISG99} and \cite{serfati98}. This area has seen substantial recent activity, mostly in the direction of extending or generalizing Marchioro's original work, see \cite{BCM00,HLN99,Iftimie999999,LN98,marchioro96,marchioro98a,marchioro98b,marchioro99,MM01,MP93}.
     
Results on confinement of vorticity are rigorous actualizations of the rough idea that single signed 2D vorticity tends to rotate around, but not to spread out. This is false if the vorticity is not single signed, which can be seen by considering the behavior of {\it vortex
pairs}, vorticity configurations that tend to translate to infinity with constant speed due to their self-induced velocity, see \cite{ISG99} for a specific smooth example. Due to the traveling wave behavior of vortex pairs, vorticity scattering in two dimensions may become complicated, and interesting, when vorticity is allowed to change sign. Recently, the authors have proved a new result on confinement of vorticity in this context. Let $\omega = \omega(t,x)$ be a solution of incompressible 2D Euler in the full plane, with compactly supported but not necessarily single signed initial data $\omega_0$ and let $M = \int \omega_0$. For any $a>0$, define the rescaling $W(t,x) \equiv t^{2a} \omega(t,t^ax)$. We have proved that if $a>1/2$ then $W(t,\cdot)$ converges weakly to $M \delta$. This means confinement, in a weak sense, of the net vorticity in a region with roughly square-root in time growth in its diameter.  
This result will appear in a forthcoming article by the authors, \cite{ILN02}.
From the point of view of scattering, this result accounts for the behavior of the net vorticity, but says very little about the behavior of vortex pairs, because these tend to be weakly self-canceling when looked at from a large spatial scale. It one wants to study vortex scattering, the relevant information is the large-time behavior of $|W(t,\cdot)|$, mainly in the case $a=1$. The present article is directed precisely at this problem, with the simplifying assumption that the vorticity be odd with respect to a straight line, single-signed on each side of the symmetry line. Another way of expressing this is to say that in this article will study the scattering of co-axial, unidirectional vortex pairs.  
    
Let $\omega = \omega(t,x)$ be the solution of the half-plane problem (\ref{vorteq}) defined for all time, associated to initial data $\omega_0$, which we assume, for simplicity, to be smooth, compactly supported and nonnegative.  A confinement result proved by Iftimie in \cite{Iftimie999999}, together with what we will prove here implies that the support of $\omega(t,\cdot)$ is contained in a rectangle of the form $(a_1 - b_1 t^{\alpha} , ct) \times (0 , a_2 + b_2 t^{\beta})$, with $a_i$ real constants, $b_i,c>0$ and $0 \leq \alpha,\beta < 1$. We wish to examine the asymptotic behavior of the vorticity on the linearly growing horizontal scale that is naturally associated with the motion of vortex pairs. The approach we use is
inspired on work on the asymptotic behavior of solutions of systems of conservation laws due to G. Q. Chen and H. Frid, see \cite{CF99}. Let $\omt (t,x) \equiv t^2 \omega(t,tx)$. The function $\omt$ has bounded $L^1$ norm and will be shown to have support in a rectangle of the form 
$(- b_1 t^{\alpha-1} , c) \times (0 , b_2 t^{\beta-1})$. Hence the family of measures $\{\omt (t,\cdot)\}_{t>0}$ is weak-$\ast$ precompact and any weak limit of subsequences of this family is of the form $\mu \otimes \delta_0$, with $\mu$ a nonnegative measure supported on the interval $[0,c]$. We will refer to such a measure $\mu$ as an {\it asymptotic velocity density}. Our main result may be stated in the following way.  

\begin{theorem} \label{mainthm1}
Suppose that the initial data $\omega_0$ for problem
(\ref{vorteq}) is such that there exists a {\bf unique} asymptotic velocity density $\mu$, i.e., $\omt (t,\cdot) \rightharpoonup \mu \otimes
\delta_0$ when $t \to \infty$. Then $\mu$ is the sum of
an at most countable set of Diracs whose supports may only accumulate
at zero. 
\end{theorem} 

The proof involves writing the PDE for the evolution of $\omt $ and using
the {\it a priori} estimates available and the structure of the
nonlinearity in a way that is characteristic of weak convergence
methods, see \cite{evans90}. We will briefly discuss the physical meaning of both the hypothesis that $\omt (t,\cdot)$ converges weakly and the conclusion regarding the structure of $\mu$. 

The study of the wavelike behavior of vortex pairs goes back to Pocklington in
\cite{pocklington}, with more recent interest going back to work of
Norbury, Deem and Zabusky and Pierrehumbert, see
\cite{dz77,norbury75,pierrehumbert80}. The existence (and abundance) of 
steady vortex pairs, which are traveling wave solution of the 2D incompressible Euler equations, i.e. vorticity shapes which propagate with constant speed without deforming, has been established in the literature in several ways, see \cite{burton88,jianfu91,norbury75}. Steady vortex pairs have been object of an extensive literature, from asymptotic studies, see \cite{yk94} and numerical studies, see \cite{pullin92} and even experimental work, see \cite{ds88}. Although some analytical results (see \cite{moffatt90}) and numerical evidence, \cite{oz82}, point to the orbital stability of steady vortex pairs under appropriate conditions, this stability is an interesting, largely open problem, see \cite{saffman95}.  

Compactly supported vortex pairs interact in a way such that the intensity of the interaction decays with the inverse of the square of the distance between them. Hence, vortex pairs moving with different speeds tend to behave like individual particles, decoupling after a large time. This is what makes the study of vortex scattering interesting in this context. Let us illustrate the point of view we want to pursue with the example of the Korteweg-deVries equation. Nonlinear scattering for the KdV is well-understood, as solutions of KdV with smooth, compactly supported initial data are expected to resolve into a scattering state composed of an $N$-soliton plus a slowly decaying dispersive tail. This fact was first formulated as a conjecture by P. Lax in 
\cite{lax68} and broadly explored through the method of inverse scattering since then. The conclusion of Theorem \ref{mainthm1} may be regarded as a weak, or averaged form of Lax's conjecture for vortex pair dynamics. Note that steady vortex pairs correspond to classical solitons in this analogy, but no existence for the multibump solutions that would be associated to the classical N-solitons has been rigorously established.  

Let us call shape space the space of smooth compactly supported vorticity configurations, identifying configurations which are related through horizontal translations. Steady vortex pairs correspond to stationary shapes with respect to Euler dynamics. There are solutions of the
two-dimensional incompressible Euler equations that describe periodic loops in shape space. Two examples of this behavior are: 1) a pair of like-signed point vortices on a half plane, which orbit one another periodically as they translate horizontally, called leapfrogging
pairs, and 2) Deem and Zabusky's translational $V$-states, which are vortex patches with discrete symmetry, see \cite{dz77}. From the point of view of scattering such solutions represent another kind of asymptotic state or, in other words, another kind of particle. Furthermore, one may well imagine solutions with quasiperiodic or chaotic behavior in shape
space. Although there is no example of either case in the literature, the passive tracer dynamics of the leapfrogging pair is known to be chaotic, see \cite{ptt95}. Possible chaotic shapes represent an interesting illustration of Theorem \ref{mainthm1}, as both the
hypothesis of weak convergence and the conclusion are clearly related to the ergodicity of shape dynamics and the self-averaging of the velocity of the center of vorticity of such generalized vortex pairs. Finally, we must mention the work of Overman and Zabusky \cite{oz82}, where they do numerical experiments on the short term scattering of pairs of translational $V$-states, the first (and only) study to date on the interaction of coaxial vortex pairs, which is the main point of the present work.      

The remainder of this article is divided into two large sections. The first one contains a discussion of confinement of vorticity in half plane flow, including two new results, horizontal confinement on the left and bounds on the distance to the boundary. The second one contains the discussion leading to our main result, together with its proof. After this second section we include a brief section with conclusions and an Appendix containing a simple illustration of half-plane vortex dynamics.

\section{Confinement of vorticity}

\subsection{Preliminary results}

Let us begin by fixing basic notation. We denote by $\Bbb{H}$ the horizontal half-plane given by $\Bbb{H} = \{ x \in \R^2 ; x_2 > 0\}$. Reflection with respect to $x_2 = 0$ will be denoted by $x = (x_1,x_2) \mapsto \wb x= (x_1, -x_2)$.  If $z = (z_1,z_2)$ then its perpendicular vector is $z^{\perp} = (-z_2,z_1)$. We use $L^p_c(\Bbb{H})$ to denote the Lebesgue space of $p$-th power integrable functions, $p \geq 1$, with compact support in $\Bbb{H}$. The dual of $L^p$ is $L^{p^{\prime}}$, with the conjugate exponent given by $p^{\prime} = p/(p-1)$. The space of bounded Radon measures is denoted by $BM$ and the Dirac delta at the origin is $\delta_0$. 

Consider the initial-boundary value problem for the incompressible 2D Euler equations in the half-plane (\ref{vorteq}) with initial vorticity $\omega_0$. If $\omega_0$ is bounded then (\ref{vorteq}) is globally well-posed since it is equivalent, through the method of images, to an initial-value problem in the full-plane, with bounded, compactly supported initial vorticity (shown to be well-posed by Yudovich in \cite{yudovich63}). The method of images consists of the observation that the Euler equations are covariant with respect to mirror-symmetry. Thus an initial vorticity which is odd with respect to reflection about the horizontal axis will remain so, and give rise to flow under which the half-plane is invariant. Conversely, the odd extension, with respect to $x_2 = 0$, of vorticity in half-plane flow gives rise to full-plane flow. This observation is especially useful in order to deduce the Biot-Savart law for half-plane flow, to recover velocity from vorticity.

Let us fix an initial vorticity $\omega_0$. We will assume 
{\em throughout this article} that $\omega_0$ is a given {\em nonnegative} function in $L^p_c(\Bbb{H})$ for some $p > 2$. If $\omega_0 \in L^p_c(\Bbb{H})$, $2 < p < \infty$, then there exists a weak solution $u$, $\omega$ of (\ref{vorteq}) associated with this initial vorticity (see \cite{LNX01}). Furthermore, $\omega(t,\cdot) \geq 0$, $t \geq 0$, and the $L^1$ and $L^p$-norms of $\omega(t,\cdot)$ are bounded by the $L^1$ and $L^p$-norms, respectively, of the initial vorticity. Using the method of images we can write the velocity $u$ in terms of vorticity $\omega$ as:
\begin{equation}\label{nBS}
u(t,x)=\int_\HH  {\Bigl[\frac{(x-y)^\perp}{2\pi|x-y|^2}-\frac{(x-\wb
  y)^\perp}{2\pi|x-\wb y|^2}\Bigr]}\omtyy\dy.  
\end{equation}
We denote the kernel appearing the integral above by:
\begin{equation} \label{theK}
 K = K(x,y) = \frac{(x-y)^\perp}{2\pi|x-y|^2}-\frac{(x-\wb
  y)^\perp}{2\pi|x-\wb y|^2},
\end{equation}
whose components are given explicitly by:
\begin{equation}\label{npK}
K_1(x,y)=\frac{y_2[y_2^2-x_2^2+(x_1-y_1)^2]}{\pi|x-\wb y|^2|x-y|^2}
\quad\text{and}\quad
K_2(x,y)=\frac{2(x_1-y_1)x_2y_2}{\pi|x-\wb y|^2|x-y|^2}.
\end{equation}

It is easy to see that 
\begin{equation}\label{K}
  |K(x,y)|\leq \frac{1}{\pi|x-y|}, 
\end{equation}
from which we can deduce the fact that, if $p >2$, then an $L^1 \cap L^p$-vorticity $\omega$ gives rise to an $L^{\infty}$-velocity $u$ with the estimate:
\[ \|u\|_{L^{\infty}(\Bbb{H})} \leq C\|\omega\|_{L^p(\Bbb{H})}^{p^{\prime}/2}\|\omega\|_{L^1(\Bbb{H})}^{1 - p^{\prime}/2}.\]
One interesting fact is that this estimate can be {\em localized}. From a technical point of view, this fact is the heart of the arguments presented in this work. To be more precise we will be using the following lemma, whose proof can be found in \cite{Iftimie999999}:

\begin{lemma}\label{sjml}
Let $a\in(0,2)$, $S\subset\R^2$ and let $h:S\rightarrow\R$, $h \geq 0$, be a function belonging to  $L^1(S)\cap L^p(S)$, $p>\frac{2}{2-a}$. Then there exists a constant $D = D_{a,p}> 0$ such that the  following estimate holds: 
$$
\int_S\frac{h(y)}{|x-y|^a}\dy
\leq D_{a,p}\|h\|_{L^p(S)}^{ap^{\prime}/2} \|h\|_{L^{1}(S)}^{1 -
  ap^{\prime}/2}, \qquad\forall x\in \R^2. 
$$   
\end{lemma}

\begin{remark}
To illustrate our use of this Lemma note that, when the function $h$ is the vorticity and $a = 1$ this estimate implies that the portion of velocity due to vorticity in a region $S$ will be small if the mass of vorticity in that region is small.   
\end{remark}

\subsection{One-sided horizontal confinement of vorticity}

Iftimie showed in \cite{Iftimie999999} that the horizontal component of the center of mass in half-plane flow travels with speed bounded below by a positive constant. This excludes any possible sublinear-in-time horizontal confinement, at least in the direction $x_1 > 0$. On the other hand, half-plane flows with nonnegative vorticity have a tendency to move to the right, resisting left ``back flow". The purpose of this section is to make this statement more precise. Our main result in this section is:

\begin{theorem}\label{thleft}
Let $\omega_0 \in L^p_c(\HH)$, $p > 2$, $\omega_0 \geq 0$. Let $u$ and $\omega$ be solutions of (\ref{vorteq}) with initial vorticity $\omega_0$. Then there exists a positive constant $D$ depending solely on the initial vorticity such that 
\begin{equation*}
\supp\omega(t,\cdot)\subset\{x\in\HH\ ;\ x_1\geq -D(t\log t)^{\frac12}\}
\end{equation*}
for all $t>2$.
\end{theorem}

Before we give the proof of Theorem \ref{thleft} we need a technical lemma, in which we obtain an estimate on the mass of vorticity in the ``back flow" region; we see that it is exponentially small.

\begin{lemma} \label{vortmassleft}
Given $k\in\N$, there exist positive constants $D_1$ and $D_2$,  
depending only on the initial vorticity and on $k$, such that
  \begin{equation*}
    \int_{y_1<-r}\omtyy\dy\leq\frac{D_1}{r^k}
  \end{equation*}
provided that $r\geq D_2(t\log t)^{\frac12}$ and $t\geq2$.
\end{lemma}

\begin{proof}
Consider the auxiliary function $\eta = \eta(s)=\frac{e^s}{1+e^s}$. It is easy to see that $\eta$ is nonnegative, increasing and
\begin{equation} \label{etaest}
|\eta^{\prime\prime} (s) | \leq \eta(s).
\end{equation} 
Set
\begin{equation*}
  f_r(t)=\int\eta\bigl(-\frac{x_1+r}{\la r}\bigr)\omtxx\dx,
\end{equation*}
where $\la>0$ will be chosen later. As $\eta$ is nonnegative and increasing we clearly have: 
\begin{equation}\label{ll2}
f_r(t)\geq\int_{x_1\leq-r}\eta \bigl(-\frac{x_1+r}{\la r}\bigr)\omtxx\dx
\geq\eta(0)\int_{x_1\leq-r}\omtxx\dx,  
\end{equation}
where we have used that for $x_1\leq-r$ we have that
$-\frac{x_1+r}{\la r}\geq0$. Therefore it suffices for our purposes to estimate $f_r(t)$.  

We will deduce a differential inequality for $f_r$ from which we estimate $f_r$. To this end we differentiate in time to find:
\[f'_r(t)=-\frac1{\la r}\int\eta'\bigl(-\frac{x_1+r}{\la
  r}\bigr)u_1(t,x)\omtxx\dx,\]
where we have used the vorticity equation (\ref{vorteq}) and integration by parts to throw derivatives onto $\eta$,
\[= -\frac1{2\pi\la r}\iint\eta'\bigl(-\frac{x_1+r}{\la
  r}\bigr)\Bigl[\frac{x_2+y_2}{|x-\wb
  y|^2}-\frac{x_2-y_2}{|x-y|^2}\Bigr]\omtxx\omtyy\dx\dy,\]
using the Biot-Savart law \eqref{nBS}, 
\[ \leq \frac1{2\pi\la r}\iint\eta'\bigl(-\frac{x_1+r}{\la
  r}\bigr)\frac{x_2-y_2}{|x-y|^2}\omtxx\omtyy\dx\dy,\]
as $\eta'$, $x_2$ and $y_2$ are positive.  Finally, we symmetrize the kernel above by making the change of variables $x\leftrightarrow y$ to obtain:
\[f'_r(t) \leq \frac1{4\pi\la r}\iint\Bigl[\eta'\bigl(-\frac{x_1+r}{\la
  r}\bigr)-\eta'\bigl(-\frac{y_1+r}{\la
  r}\bigr)\Bigr]\frac{x_2-y_2}{|x-y|^2}\omtxx\omtyy\dx\dy\]
\[ \leq \frac1{4\pi\la r}\iint\frac{|x_1-y_1|}{\la r}|\eta''(\theta_{x,y})|\frac{|x_2-y_2|}{|x-y|^2}\omtxx\omtyy\dx\dy,\]
by the mean value theorem, with $\theta_{x,y}$ some point between 
$-\frac{x_1+r}{\la r}$ and $-\frac{y_1+r}{\la r}$. 

Next we use (\ref{etaest}) and the fact that $\eta$ is nonnegative and increasing to deduce that 
\begin{equation*}
|\eta''(\theta_{x,y})|\leq |\eta(\theta_{x,y})|  
\leq \eta\bigl(-\frac{x_1+r}{\la r}\bigr)
+\eta\bigl(-\frac{y_1+r}{\la r}\bigr).
\end{equation*}
Since $|x_1-y_1|\ |x_2-y_2|\leq  |x-y|^2$ we finally obtain the differential inequality: 
\begin{equation*}
  f'_r(t)
\leq\frac 1{4\pi\la^2 r^2}\iint \Bigl[\eta\bigl(-\frac{x_1+r}{\la r}\bigr)
+\eta\bigl(-\frac{y_1+r}{\la r}\bigr)\Bigr]\omtxx\omtyy\dx\dy
= \frac{\|\omega_0\|_{L^1}}{2\pi\la^2 r^2}\,f_r(t), 
\end{equation*}
where we have used that the $L^1$-norm of $\omega(t,\cdot)$ is constant
in time. Integration now yields
\begin{equation*}
f_r(t)\leq f_r(0)\exp\Bigl(t\frac{\|\omega_0\|_{L^1}}{2\pi\la^2 r^2}\Bigr).  
\end{equation*}
Clearly we may assume, without loss of generality, that
$\supp\omega_0\subset\{x_1\geq0\}$. Then
\begin{equation*}
f_r(0)=\int\eta\bigl(-\frac{x_1+r}{\la r}\bigr)\omega_0(x)\dx
\leq \eta\bigl(-\frac1\la\bigr)\|\omega_0\|_{L^1}
\leq \exp\bigl(-\frac1\la\bigr)\|\omega_0\|_{L^1}. 
\end{equation*}
Hence, we infer that 
\begin{equation*}
f_r(t)\leq \|\omega_0\|_{L^1} \exp\Bigl(t\frac{\|\omega_0\|_{L^1}}{2\pi\la^2 r^2}-\frac1\la\Bigr).  
\end{equation*}
In view of \eqref{ll2}, to finish the proof it is now sufficient to
choose $\la$ such that
\begin{equation*}
\exp\Bigl(t\frac{\|\omega_0\|_{L^1}}{2\pi\la^2 r^2}-\frac1\la\Bigr)
\leq \frac1{r^k}=\exp(-k\log r).  
\end{equation*}
The choice 
\begin{equation*}
\la=\frac1{2k\log r}  
\end{equation*}
is convenient provided that the following inequality holds
\begin{equation} \label{maldicao}
\frac{r^2}{\log r}\geq t\frac{2k\|\omega_0\|_{L^1}}{\pi}.  
\end{equation}
Notice that the function $r \mapsto r^2/\log r$ is nondecreasing if $r>e$. Hence, choosing $D_2$ sufficiently large, it is easy to ensure (\ref{maldicao}) if $r \geq D_2(t\log t)^{\frac12}$ and $t\geq2$. This completes the proof.
\end{proof}

Next we use Lemma \ref{vortmassleft} to estimate the horizontal velocity. 

\begin{proposition}\label{propleft}
Under the hypothesis of Theorem \ref{thleft}, there exist positive constants $D_3$ and $D_4$ such that
\begin{equation*}
|u_1(t,x)|\leq\frac{D_3}{|x_1|}
\quad\text{for all}\quad t\geq2\text{ and }x\in\HH\text{ such that }
x_1\leq -D_4(t\log t)^{\frac12}.  
\end{equation*}
\end{proposition}

\begin{proof}
 We will estimate directly $u_1(t,x)$. From the Biot-Savart law
\eqref{nBS} and the decay estimate \eqref{K} it follows that
\begin{align*}
|u_1(t,x)|
&\leq \int\frac1{\pi|x-y|}\omtyy\dy\\
&\leq \int_{y_1<x_1/2}\frac1{\pi|x-y|}\omtyy\dy  
+\int_{y_1 \geq x_1/2}\frac1{\pi|x-y|}\omtyy\dy\\
&\leq \frac{D_{1,p}}\pi \|\omega_0\|_{L^p}^{p^{\prime}/2}
\Bigl(\int_{y_1<x_1/2}\omtyy\dy\Bigr)^{1-p^{\prime}/2}
+\frac 2{\pi |x_1|}\|\omega_0\|_{L_1},
\end{align*}
using Lemma \ref{sjml} with $a=1$. We have also used that both the $L^1$ and the $L^p$-norms of $\omega(t,\cdot)$ are bounded by their initial values.

Let 
\[k = \left[\frac{2}{2-p^{\prime}}\right] + 1,\]
where $[a]$ denotes the largest integer smaller than $a$. Choose $D_2$  as in Lemma \ref{vortmassleft} and let $x$ satisfy 
$x_1 \leq -D_4(t\log t)^{\frac12}$ with $D_4 = 2D_2$. The conclusion then follows from Lemma \ref{vortmassleft} with $D_3$ computed accordingly.
\end{proof}

We will finish this section with the proof of the horizontal confinement to the left. 

\begin{proof}[Proof of Theorem \ref{thleft}]
Let $D_3$ and $D_4$ be as in Proposition \ref{propleft}. 
If need be increase the values of $D_3$ and $D_4$ to show that any trajectory which reaches the region $\{x_1\leq -D_4(t\log t)^{\frac12}\}$ does not have enough horizontal velocity to go past the line $x_1= -2D_4(t\log t)^{\frac12}$. This proves that every trajectory lies in the
region $\{x_1\geq -2D_4(t\log t)^{\frac12}\}$ (with $D_4$ depending 
on the initial position of the trajectory); in particular, the support of the evolved vorticity stays in that region.
\end{proof}

\subsection{Proximity to the boundary}

For smooth flows it is easy to see that, if the support of the initial vorticity lies in the interior of the half-plane, then the support never reaches the boundary due to uniqueness of solutions of ODEs. Even if the flow is not smooth this remains accurate when understood in the context of the R. DiPerna and P.-L. Lions' theory of ODEs in Sobolev spaces \cite{DL89}. However, we are left with no information on how far a particle path must remain from the boundary. The purpose of this section is to examine this issue. We will show that particle paths must stay at least as far as $e^{-C_1 e^{C_2t}}$ away from the boundary, for some positive constants $C_1$, $C_2$. This result is interesting from the point of view of confinement, but it will not be used for the development of large time asymptotics.  
A similar concern was addressed by N. Depauw in his work on vortex patches in a bounded domain, see \cite{depauw99}. 

We will use throughout this section the notation $\logt$ to denote the map $s \mapsto \logt (s)  =\sqrt{1+\log^2(s)}$. Note that $\bigl|\dert\logt t\bigr|\leq\frac1t$ for all $t>0$. We denote a particle path by $X=X(t)$, so that 
\[ \frac{d}{dt}X = u(t,X).\]

\begin{theorem}\label{d1}
Assume that $\omega_0\in L^1\cap L^\infty$.
Then there exists a positive constant $D$, depending only on the initial vorticity, such that every trajectory $X$ verifies
\begin{equation*}
X_2(t)\geq e^{-\logt(X_2(0))e^{Dt}}  
\end{equation*}
for all times $t\geq 0$.
\end{theorem}

\begin{remark}
If the initial vorticity is compactly supported in the half-plane then, as vorticity is transported by the flow, it follows that there exists a positive constant $D_1$, depending only on the initial vorticity, such that
\begin{equation*}
\supp\omega(t,\cdot)\subset\{x\in\HH\ ;\ x_2\geq e^{-D_1e^{D_1t}}\}
\end{equation*}
for all times $t\geq 0$.
\end{remark}

\begin{proof}[Proof of Theorem \ref{d1}]
We will prove that for all $x\in\HH$, one has that
\begin{equation}
  \label{ss1}
  |u_2(t,x)|\leq Dx_2\logt x_2
\end{equation}
for some constant $D>0$ depending only on the initial vorticity. 
Assuming that this is true, if $X=(X_1,X_2)$ is a particle trajectory then:  
\begin{equation*}
|X_2'|\leq DX_2\logt X_2,  
\end{equation*}
which in turn implies that
\begin{equation*}
 \Bigl| \dert\log\logt X_2\Bigr|=\Bigl|\frac1{\logt X_2}X_2'\logt'(X_2)\Bigr|
\leq\frac{|X_2'|}{X_2\logt X_2}\leq D.
\end{equation*}
After integration, we infer that
\begin{equation*}
\log\logt X_2\leq Dt+\log\logt(X_2(0)),  
\end{equation*}
that is,
\begin{equation*}
\logt X_2\leq\logt(X_2(0)) e^{Dt}.  
\end{equation*}
We therefore find that
\begin{equation*}
  -\log X_2\leq \logt(X_2(0)) e^{Dt},
\end{equation*}
so that
\begin{equation*}
X_2(t)\geq e^{-\logt(X_2(0))e^{Dt}},  
\end{equation*}
which proves Theorem \ref{d1}.

\bigskip

Let us turn to the proof of \eqref{ss1}. Fix $x\in \HH$ and start by noting that, 
by the Biot-Savart law \eqref{nBS} and \eqref{npK},
\begin{equation*}
u_2(t,x)=\frac{2x_2}{\pi}\int \frac{(x_1-y_1)y_2}
{|x-y|^2|\wb x-y|^2}\omtyy\dy \equiv \frac{2x_2}{\pi}\int L_1(x,y)\omtyy\dy.  
\end{equation*}

We begin by observing that $|x_1 - y_1| \leq |x-y|$ and, since $x_2$, $y_2 > 0$, $|y_2| \leq |x_2 + y_2| \leq |\wb x-y|$, so that
\[|L_1(x,y)| \leq \frac{1}{|x-y||\wb x-y|}.\]
It can be easily checked that
\[ x_2 + y_2 \geq \frac{|x_2 - y_2| + x_2}{2}.\]
Using this estimate we find:
\[|\wb x-y| \geq \frac{1}{2}\left( |x_1 - y_1|+|x_2 + y_2| \right) \geq \frac{|x_1 - y_1| + 
|x_2 - y_2| + x_2}{4} \geq \frac{|x-y| + x_2}{4}. \]
Finally, we obtain the following estimate for $L_1$:
\begin{equation} \label{L1est}
|L_1(x,y)| \leq \frac{4}{|x-y|(|x-y| + x_2)}.
\end{equation}

In order to estimate $u_2(t,x)$ we first estimate the contribution of vorticity-bearing particles far from $x$: 
\begin{equation} \label{farfield}
\int_{ \{ |y-x| \geq 1 \} }|L_1(x,y)|\omtyy\dy\leq 
4\int\omtyy\dy\leq 4\|\omega_0\|_{L^1}. 
\end{equation}
From  (\ref{L1est}) and  (\ref{farfield}) we  deduce that 
\begin{equation} \label{u2est}
|u_2(t,x)|\leq Cx_2+ C x_2\int_{ \{ |x-y|\leq 1 \}} 
\frac{1}{|x-y|(|x-y| + x_2)}\omtyy\dy
\end{equation}
for some constant $C>0$. Changing to polar coordinates and estimating $\omega(t,\cdot)$ by  the $L^{\infty}$-norm of $\omega_0$, it follows that
\begin{equation*}
|u_2(t,x)|\leq Cx_2+ C x_2\int_0^1 \frac{1}{r(r + x_2)}r \,\text{d}r
=Cx_2+Cx_2(\log(1+x_2)-\log x_2),
\end{equation*}
for some constant $C$. Relation \eqref{ss1} now follows.
\end{proof}

\begin{remark}
We call attention to the fact that this result holds without any condition on the sign of vorticity. 
\end{remark}

\subsection{Vertical confinement}

Another piece of information on confinement of vorticity for half-plane
flows stems from the conservation of the second component of the center
of vorticity. Such a result was obtained by Iftimie in
\cite{Iftimie999999} (see Theorem 3 and also Remark 3 of
\cite{Iftimie999999}). The resulting  estimate will be used in what follows, so that we include its precise statement here for the sake of completeness. 
 
\begin{theorem}[\cite{Iftimie999999}]\label{thup}
If the initial vorticity $\omega_0$ belongs to $L^p_c(\HH)$, $p>2$, then 
there exists a constant $D>0$, depending solely on $\omega_0$ and $p$, such that 
\begin{equation*}
\supp\omega(t,\cdot)\subset\{x\in\HH\ ;\ x_2\leq D(t\log t)^{\frac13}\}
\end{equation*}
for all $t>2$, where $\omega(t,\cdot)$ is a weak solution of (\ref{vorteq}) having initial vorticity $\omega_0$.
\end{theorem}

The complete proof can be found in \cite{Iftimie999999}.
 
\section{Asymptotic behavior of nonnegative vorticity in the half-plane} 

We now turn to our main concern in this paper, the rigorous study of the asymptotic behavior of flows with nonnegative vorticity in the half-plane. We divide this section in three subsections. In the first one we introduce the self-similar rescaling of the flow which encodes the scattering information we wish to study, we write an evolution equation for the rescaled vorticity and we interpret the vortex confinement information obtained in the previous section in terms of the new scaling. The second subsection is the technical heart of this article, where we study the behavior of the nonlinearity in the equations with respect to the self-similar scaling. Finally, in the third subsection we use the information obtained to prove our main result.      

\subsection{Rescaled vorticity and asymptotic densities}

        One key feature of vortex dynamics in a half-plane is nonlinear wave propagation. In order to examine wave propagation it is natural to focus on a self-similar rescaling of physical space, as has been performed by Chen and Frid in the context of systems of conservation laws, see \cite{CF99}. Let us fix, throughout this section, a nonnegative function $\omega_0\in L^p_c(\HH)$, $p>2$, and $\omega = \omega(t,\cdot)$, $u = u(t,\cdot)$, solutions of (\ref{vorteq}) with initial vorticity $\omega_0$. Set  
\begin{equation} \label{rescaling}
\omty=t^2\omega(t,ty)\quad\text{and}\quad\wt u(t,y)=tu(t,ty), 
\end{equation}
the rescaled vorticity and velocity, respectively. The scaling above respects the elliptic system relating velocity and vorticity so that we still have
\[\left\{\begin{array}{l}
\dive\wt u=0 \\ 
\curl\wt u=\omt. \end{array} \right.\]
It is immediate that $\wt u_2(t,x_1,0)=0$ and therefore we can recover $\wt u$ from $\omt$ by means of the Biot-Savart law for the half-plane: 
\begin{equation}\label{BS}
\wt u(t,x)=\int_\HH K(x,y)\omty\dy,
\end{equation}
with $K$ defined in \eqref{theK}. 

Let $M = \| u \|_{L^{\infty}(\R_+\times\HH)}$. Then the confinement estimates for vorticity in the half-plane, in particular Theorems \ref{thleft} and \ref{thup} and the fact that the vorticity $\omega$ is transported by the velocity $u$, 
imply that there exists a constant $C>0$ such that:

\begin{equation*}
\supp\omega(t,\cdot)\subset\bigl[-C(t\log t)^{\frac12},C_0+Mt\bigr]
\times\bigl[0,C(t\log t)^{\frac13}\bigr]\quad\text{for all }t\geq2,  
\end{equation*}
where $C_0=\sup\{x_1\ ;\ x\in\supp\omega_0\}$.
This in turn implies a sharp asymptotic localization on 
$\supp\omt(t,\cdot)=\frac1t\supp\omega(t,\cdot)$, namely:

\begin{equation} \label{suppomt}
\supp\omt(t,\cdot)\subset\Bigl[-C\bigl(\frac{\log t}{t}\bigr)^{\frac12},\frac{C_0}t+M\Bigr]
\times\Bigl[0,C\bigl(\frac{\log t}{t^2}\bigr)^{\frac13}\Bigr].    
\end{equation}

Next, from the vorticity equation one may derive a transport equation for the evolution of $\omty$, which takes the form:

\begin{equation} \label{1}
\partial_t\omty-\frac1t\dive\bigl[y\omty\bigr]+\frac1{t^2}\dive\bigl[\wt u(t,y)\omty\bigr]=0.  
\end{equation}

Using the scaling (\ref{rescaling}) we find
\begin{equation}\label{a3}
\nlp {q}{\omt(t,\cdot)}=t^{2(1-\frac1q)}\nlp {q}{\omega(t,\cdot)}
\leq t^{2(1-\frac1q)}\nlp {q}{\omega_0}\quad\forall q\in[1,p].  
\end{equation}

Furthermore, the $L^1$-norm of $\omt$ is conserved in time.
We wish to treat $\omt$ as a bounded $L^1$-valued function of time, possessing nonnegative measures as weak-$\ast$ limits for large time. The confinement estimate (\ref{suppomt}) implies that any weak-$\ast$ limit of $\omt$ must have the structure $\mu \otimes \delta_0(x_2)$, with the support of $\mu$ contained in the interval $[0,M]$.   

It is in the nature of the self-similar rescaling (\ref{rescaling}) that much of the scattering behavior of the flow is encoded in the measure $\mu$. This measure is the main subject of the remainder of this article, and, as such, deserves an appropriate name.

\begin{definition} \label{asympveldensty}
Let $\mu \in BM([0,M])$ be a nonnegative measure such that there exists a sequence of times $t_k \to \infty$ for which 
\[ \omt (t_k,\cdot) \rightharpoonup \mu \otimes \delta_0 \mbox{ in the weak-$\ast$ topology of bounded measures, as } t_k \to \infty.\]
Then we call $\mu$ an {\bf asymptotic velocity density} associated to $\omega_0$.
\end{definition}

It can be readily checked that, if $\omega (t,x) = \omega_0 (x_1 -\sigma t,x_2)$, then there exists a unique asymptotic velocity density $\mu$, which is a Dirac delta at position $(\sigma,0)$ with mass given by the integral of $\omega_0$. For a general flow an asymptotic velocity density encodes information on typical velocities with which different portions of vorticity are traveling. 

\subsection{The key estimate}

Our purpose in this article is to understand the structure of the asymptotic velocity densities. To do so we make use of the evolution equation (\ref{1}) for $\widetilde{\omega}$  and we examine   the behavior for large time of each of its terms. The main difficulty in doing so is understanding the behavior of the nonlinear term 
$\mbox{ div }(\widetilde{u} \ \widetilde{\omega})$, which is our goal in this subsection. 

We begin with two general measure-theoretical lemmas which will be needed in what follows. These are standard exercises in real analysis and we include the proofs only for the sake of completeness. Recall that a measure is called {\it continuous} if it attaches zero mass to points.

\begin{lemma}\label{l1}
Let $\mu$ be a finite and compactly supported nonnegative measure on $\Bbb{R}$. Then $\mu$ is the sum of a nonnegative continuous measure $\nu$ and a countable sum of positive Dirac measures (the discrete part of $\mu$). Moreover, for every $\ep>0$ there exists
$\delta>0$ such that, if $I$ is an interval of length less than $\delta$,
then $\nu(I)\leq\ep$.  
\end{lemma}

\begin{proof}
Let $A=\{x\ ;\ \mu(\{x\})\neq 0\}$. Then $A$ is countable; indeed, $A=\bigcup_n
A_n$ and each $A_n=\{x\ ;\ \mu(\{x\})\geq 1/n\}$ must be finite because $\mu$ is
finite. Hence we may write $A=\{x_1,x_2,\dots\}$ and $m_j=\mu(\{x_j\})$. 
Of course, $\nu=\mu-\sum_j m_j \delta_{x_j}$ is a continuous, nonnegative measure.

Let $J$ be a compact interval containing the support of $\mu$. For each 
$x\in J$, it follows that, since $\nu(\{x\})=0$, there exists $\delta_x>0$ such that
$\nu([x-\delta_x,x+\delta_x])\leq \ep/2$. Let
$J_x=[x-\delta_x,x+\delta_x]$. Then $J\subset\bigcup_{x\in J}J_x$ so that, using the fact that $J$ is compact, we can extract a finite subcover, $J\subset J_{x_1}\cup\dots\cup
J_{x_n}$. The interval $J$ is now divided in a finite number of disjoint
intervals (not necessarily $J_{x_1}, J_{x_2},\dots,J_{x_n}$), each having $\nu$--measure less than $\ep/2$. It is then sufficient to choose $\delta$ equal to one half of the minimum length of these intervals. For that choice of $\delta$, it is clear that an interval of
length $\delta$ cannot intersect more than two of the disjoint intervals
constructed above so that its $\nu$--measure will be less than $\ep$.
\end{proof}

\begin{lemma}\label{l-1}
Let $\gamma_n$ be a sequence of nonnegative Radon measures on $\wb\HH$, converging weakly
to some measure $\gamma$, and having the supports uniformly bounded in the vertical
direction. Then, for every compact interval $[a,b]$ one has that
\begin{equation*}
\limsup_{n\to\infty}\gamma_n([a,b]\times\R_+)\leq\gamma([a,b]\times\R_+).  
\end{equation*}
\end{lemma}

\begin{proof}
Fix $\ep>0$. Since
\begin{equation*}
\gamma([a,b]\times\R_+)=\lim_{\delta\to0}\gamma([a-\delta,b+\delta]\times\R_+),  
\end{equation*}
there exists $\delta>0$ such that
\begin{equation*}
\gamma([a-\delta,b+\delta]\times\R_+)< \gamma([a,b]\times\R_+)+\ep. 
\end{equation*}

Let $\phi$ be a continuous function supported in $(a-\delta,b+\delta)$
and such that $0\leq\phi\leq1$ and $\phi\bigl|_{[a,b]}=1$. According to
the hypothesis, we have that
$\scal{\gamma_n(y)}{\phi(y_1)}\to\scal{\gamma(y)}{\phi(y_1)}$, so there exists $N$ such that
\begin{equation*}
\scal{\gamma_n(y)}{\phi(y_1)}\leq\scal{\gamma(y)}{\phi(y_1)}+\ep\quad\forall n\geq N.  
\end{equation*}

From the hypothesis on the test function $\phi$ it follows that, for all
$n\geq N$,
\begin{multline*}
\gamma_n([a,b]\times\R_+)\leq\scal{\gamma_n(y)}{\phi(y_1)}\leq\scal{\gamma(y)}{\phi(y_1)}+\ep\\
\leq\gamma([a-\delta,b+\delta]\times\R_+)+\ep\leq\gamma([a,b]\times\R_+)+2\ep.   
\end{multline*}
We deduce that 
\begin{equation*}
\limsup_{n\to\infty}\gamma_n([a,b]\times\R_+)\leq\gamma([a,b]\times\R_+)+2\ep.  
\end{equation*}
The desired conclusion follows by letting $\ep\to0$.
\end{proof}

Let us now return to the study of the asymptotic behavior of vorticity. Let $\omega_0 \geq 0$ be a fixed function in $L^p_c(\HH)$, for some $p> 2$, and let $u$, $\omega$ be solutions of (\ref{vorteq}), with $\widetilde{u}$, $\widetilde{\omega}$ defined in (\ref{rescaling}). Let  $\mu$ be an asymptotic velocity density associated to $\omega_0$. Then $\mu$ is a nonnegative measure in $BM([0,M])$, with $M = \|u\|_{L^{\infty}}$, and by Lemma \ref{l1}, $\mu$ can be written as

\begin{equation} \label{mudecomp}
\mu = \nu + \sum_{i=1}^{\infty} m_i \delta_{\alpha_i},
\end{equation}
where $\nu$ is the continuous part of $\mu$ and $\alpha_i \in [0,M]$. As $\omega_0 \geq 0$ it follows that $m_i \geq 0$ and, as $\mu$ is a bounded measure, 
$\sum_{i=1}^{\infty} m_i < \infty$. Furthermore we can assume without loss of generality that $\alpha_i \neq \alpha_j$ in the decomposition (\ref{mudecomp}).
  
Let $\{t_k\}$ be a sequence of times approaching infinity such that 
\[ \wt \omega(t_k,\cdot) \rightharpoonup \mu \otimes \delta_0(x_2),\] 
as $k\to\infty$, weak-$\ast$ in $BM(\overline\HH)$. The following proposition is what we refer to as the key estimate in the title of this subsection.

\begin{proposition}\label{nonlinterm} 
Let $\psi \in C^0(\R)$. Then there exists a constant $D>0$, depending only on $p$, such that the following estimate holds: 
\begin{equation}
  \label{p31}
\limsup_{k \to \infty} 
\left| \int_{\HH} \psi(y_1)\frac{\wt u_1 (t_k,y)}{t_k}\wt\omega(t_k,y)dy
\right| \leq D \nlp p{\omega_0}^{\frac{p'} 2} \sum_{i=1}^{\infty} m_i^{2-\frac{p'}{2}} | \psi(\alpha_i) |. 
\end{equation}
\end{proposition}
\begin{remark}
It will be clear from the proof that the constant $D$ can be chosen as
 $D=D_{1,p}\pi^{-1}$, where $D_{1,p}$ is
the constant of Lemma \ref{sjml}.  
\end{remark}

Before giving the proof of Proposition \ref{nonlinterm}, let us motivate
the statement with the following example. Consider a steady vortex pair
with vorticity given by $\omega(t,x)=\omega_0(x_1-\sigma t,x_2)$ and
velocity $u(t,x)=u_0(x_1-\sigma t,x_2)$. Then it is easy to see that the
rescaled nonlinear term $\frac{\wt u_1}{t}\omt$ converges to $\sigma
m\delta_\sigma\otimes\delta_0$ where $m=\int\omega_0\dx$. Based on this example, one
would expect the right-hand side of \eqref{p31} to be $\sum_i \al_i m_i
|\psi(\al_i)|$ instead. On the other hand, for the steady vortex pair, it
can be easily checked that 
\begin{equation*}
  \sigma=\frac{1}{\int \omega_0\dx}\int (u_0)_1\omega_0 \dx \leq \nlp\infty{u_0} .
\end{equation*}
Using Lemma \ref{sjml} we infer that
\begin{equation*}
 |\sigma|\leq D \nlp p{\omega_0}^{p'/2} m^{1-p'/2}. 
\end{equation*}
which then implies that, as measures, the weak limit of $\frac{\wt
  u_1}{t}\omt$ is less than $D \nlp p{\omega_0}^{p'/2}
  m^{2-p'/2}\delta_\sigma\otimes\delta_0$. Hence, in light of this example we see that estimate \eqref{p31} is weaker than what might be expected, but nevertheless it is consistent with the behavior of steady vortex pairs. 

\begin{proof}[Proof of Proposition \ref{nonlinterm}]
Let us denote the integral we wish to estimate by $B_k$, so that
\begin{equation} \label{bk}
 B_k \equiv \int_{\HH} \psi(y_1)\frac{\wt u_1 (t_k,y)}{t_k}\wt\omega(t_k,y)dy.
\end{equation}

Fix $\ep>0$ throughout. Since  $\sum_{i=1}^\infty m_i<\infty$  there exists $N=N(\ep)$ such that
\begin{equation*}
  \sum_{i>N}m_i<\frac\ep4.
\end{equation*}

Additionally, it is easy to find $\delta = \delta(\ep) > 0$ such that, if $I$ is an interval, 
$ |I| \leq \delta$, then 
\begin{equation}\label{a13}
\nu(I)< \frac{\ep}{4},
\end{equation}
by using Lemma \ref{l1}, and also  
\begin{equation}\label{a12}
\mu([\al_i-2\delta,\al_i+2\delta])<m_i(1+\ep),\quad i=1,\dots,N,
\end{equation}
\begin{equation} \label{a1}
[\al_i-\delta,\al_i+\delta]\cap [\al_j-\delta,\al_j+\delta]=\emptyset,
\quad i\neq j\in\{1,\dots,N\},
\end{equation}
\begin{equation} \label{a2}
|\psi(y_1)-\psi(\al_i)|<\ep \quad \forall \ y_1 \in 
[\al_i-\delta,\al_i+\delta],\ i=1,\dots,N.  
\end{equation}

In view of Lemma \ref{l-1} and relation \eqref{a12}, there exists $K_0$ such that, if $k > K_0$ then 
\begin{equation}
  \label{a4}
\int\limits_{[\al_i-2\delta,\al_i+2\delta]\times\R_+}\wt\omega(t_k,y)\dy
<m_i(1+\ep)\qquad\forall i=1,\dots,N.  
\end{equation}

Consider now an interval
$I\subset\R\setminus\bigcup\limits_{i=1}^N(\al_i-\frac\delta2,\al_i+\frac\delta2)$
of length at most $\delta$. According to relation \eqref{a13}
\begin{equation*}
  \nu(I)<\frac\ep4.
\end{equation*}
On the other hand $\mu-\nu$, the discrete part of $\mu$, restricted
to $I$ avoids the Diracs at $\al_1,\dots,\al_N$ so that
\begin{equation*}
  (\mu-\nu)(I)\leq\sum_{i>N}m_i<\frac\ep4.
\end{equation*}

Therefore
\begin{equation}
  \label{a14}
\mu(I)< \frac\ep2.  
\end{equation}

Given a compact interval $\mathcal{J} \subset\R\setminus\bigcup\limits_{i=1}^N(\al_i-\frac\delta2,\al_i+\frac\delta2)$  of length at most $\delta$ we can use (\ref{a14}) and Lemma \ref{l-1} together with the fact that $\omt(t_k,\cdot)\rightharpoonup\mu\otimes\delta$ to find $K_0$ large enough so that, in addition to (\ref{a4}), we have 
\[ \int_{\mathcal{J}\times \R_+} \widetilde{\omega}(t_k,y)\dy < \frac{\varepsilon}{2},\]
for any $k > K_0$. We wish to show that this $K_0$ can be chosen {\em independently} of $\mathcal{J}$, but we shall have to pay a price, namely the estimate above will hold with $\ep$ on the right-hand-side, instead of $\ep /2$.

Let $J$ be a compact interval such that $J\times\R_+$ contains the
support of $\omt(t,\cdot)$ for all $t$. We write the set $J\setminus\bigcup\limits_{i=1}^N(\al_i-\frac\delta2,\al_i+\frac\delta2)$ as a finite disjoint union of intervals $I_j$, each of which we subdivide into intervals of length exactly $\delta$, together with an interval of size at most $\delta$, this being the right-most subinterval of $I_j$. This way the set
$J\setminus\bigcup\limits_{i=1}^N(\al_i-\frac\delta2,\al_i+\frac\delta2)$
can be written as the union of intervals $J_1,\dots,J_l$ of length precisely $\delta$ plus some remaining intervals $J_{l+1},\dots,J_{L}$ of length strictly less than $\delta$. According to \eqref{a14}, we have that
\begin{equation*}
\mu(J_i)<\frac\ep2\qquad\forall i=1,\dots,L.  
\end{equation*}
Next we apply Lemma \ref{l-1} and use the fact that
$\omt(t,\cdot)\rightharpoonup\mu\otimes\delta$, to obtain $K_0$ such that (\ref{a4}) is satisfied together with: 
\begin{equation}
  \label{a15}
\int\limits_{J_i\times\R_+}\widetilde{\omega}(t_k,y)\dy<\frac\ep2
\qquad\forall i=1,\dots, L,\quad k > K_0.  
\end{equation}

Let $I$ be a subinterval of
$\R\setminus\bigcup\limits_{i=1}^N(\al_i-\frac\delta2,\al_i+\frac\delta2)$
of length less than $\delta$. It is easy to see that $I$ can intersect at most two of the intervals $J_i$ as otherwise, by construction, this would imply it had to contain an interval of length precisely $\delta$. According to \eqref{a15} we deduce that $\int_{I\times\R_+}\widetilde{\omega} (t_k,y)\dy<\ep$ for all $k>K_0$. We have just shown that, if $I$ is an interval of length at most $\delta$, $I \subset 
\R\setminus\bigcup\limits_{i=1}^N(\al_i-\frac\delta2,\al_i+\frac\delta2)$ then 
\begin{equation} \label{a5}
\int\limits_{I\times\R_+}\widetilde{\omega}(t_k,y)\dy<\ep,\ \  \forall k > K_0.
\end{equation}

\medskip

Let $k > K_0$ and  set
\begin{equation*}
E_i=[\al_i-\delta,\al_i+\delta]\times\R_+,\quad
F_i=[\al_i-2\delta,\al_i+2\delta]\times\R_+,\quad
E=E_1\cup\dots\cup E_N.  
\end{equation*}

According to \eqref{a1}, the sets $E_1,\dots,E_N$ are disjoint, so we
can write $B_k$, defined in (\ref{bk}), as:
\[B_k= \sub{B_{k1}}{\sum_{i=1}^N\int_{E_i}\psi(y_1)\frac{\wt u_1(t_k,y)}{t_k}
\wt\omega(t_k,y)\dy} 
+\sub{B_{k2}}{\int_{E^c}\psi(y_1)\frac{\wt u_1(t_k,y)}{t_k}\wt\omega(t_k,y)\dy}. \]
  
We will estimate separately $B_{k1}$ and $B_{k2}$. Note that both estimates rely in an essential way on the Biot-Savart law and the fact that the kernel can be estimated by $|x-y|^{-1}$ (see (\ref{K})). In the remainder of this proof we will denote by $C$ a constant which is independent of $\vare$ and $t$. 

\paragraf{Estimate of $B_{k1}$} Using the Biot-Savart law \eqref{BS} and relation \eqref{K}, one can bound $B_{k1}$ as follows:
{\allowdisplaybreaks
\begin{align*}
|B_{k1}|
&\leq \sum_{i=1}^N \iint\limits_{\substack{x\in\HH\\ y\in
    E_i}}\frac{|\psi(y_1)|}{\pi|x-y|}\frac{\wt\omega(t_k,x)}{t_k}\wt\omega(t_k,y)\dx\dy\\    
&=\frac{1}{t_k}
\sum_{i=1}^N \iint\limits_{\substack{|x-y|\geq \delta\\ x\in\HH,y\in
    E_i}}\frac{|\psi(y_1)|}{\pi|x-y|}\wt\omega(t_k,x)\wt\omega(t_k,y)\dx\dy\\
&\ \hskip5cm + \frac{1}{t_k} \sum_{i=1}^N \iint\limits_{\substack{|x-y|<\delta\\ x\in\HH,y\in
    E_i}}\frac{|\psi(y_1)|}{\pi|x-y|}\wt\omega(t_k,x)\wt\omega(t_k,y)\dx\dy\\
&\leq \frac{\sup|\psi|}{\pi t_k \delta}\nlp1\omt\sum_{i=1}^N\int_{E_i}\omt
+
\frac{1}{t_k}\sum_{i=1}^N \iint\limits_{\substack{|x-y|<\delta\\ x\in\HH,y\in
    E_i}}\frac{|\psi(y_1)|}{\pi|x-y|}\wt\omega(t_k,x)\wt\omega(t_k,y)\dx\dy\\
&\leq \frac{C}{\delta t_k}
+\sum_{i=1}^N \iint\limits_{\substack{|x-y|<\delta\\ x\in\HH,y\in
    E_i}}\frac{|\psi(y_1)|}{\pi t_k |x-y|}\wt\omega(t_k,x)\wt\omega(t_k,y)\dx\dy.
\end{align*}
}

According to \eqref{a2}, for $y\in E_i$ we have that
$|\psi(y_1)-\psi(\al_i)|<\ep$. We therefore deduce that
\begin{equation*}
|B_{k1}|\leq \frac{C}{\delta t_k}
+\sum_{i=1}^N \frac{|\psi(\al_i)|+\ep}{\pi t_k}\iint\limits_{\substack{|x-y|<\delta\\ x\in\HH,y\in
    E_i}}\frac1{|x-y|}\wt\omega(t_k,x)\wt\omega(t_k,y)\dx\dy.  
\end{equation*}

Applying Lemma \ref{sjml} yields
\begin{multline*}
\iint\limits_{\substack{|x-y|<\delta\\ x\in\HH,y\in E_i}}
          \frac1{|x-y|}\wt\omega(t_k,x)\wt\omega(t_k,y)\dx\dy 
\leq\int_{E_i}\Bigl(\int\limits_{[y_1-\delta,y_1+\delta]\times\R_+}
          \frac{\wt\omega(t_k,x)}{|x-y|}\dx\Bigr)\wt\omega(t_k,y) \dy\\
\leq D_{1,p}\int_{E_i}\Bigl(\int\limits_{[y_1-\delta,y_1+\delta]\times\R_+}\wt\omega(t_k,x)\dx\Bigr)^{1-\frac{p'}2}\nlp
          p\omt^{\frac{p'}2} \ \wt\omega(t_k,y)\dy.
\end{multline*}

Now, if $y\in E_i$ then $[y_1-\delta,y_1+\delta]\subset[\al_i-2\delta,\al_i+2\delta]$, so that
$[y_1-\delta,y_1+\delta]\times\R_+\subset F_i$. Hence 
\begin{align*}
\iint\limits_{\substack{|x-y|<\delta\\ x\in\HH,y\in E_i}}
          \frac1{|x-y|}\wt\omega(t_k,x)\wt\omega(t_k,y)\dx\dy 
&\leq D_{1,p} \Bigl(\int_{F_i}\wt\omega(t_k,x)\dx\Bigr)^{1-\frac{p'}2}
t_k\nlp p{\omega_0}^{\frac{p'}2} \left( \int_{E_i}\wt\omega(t_k,y) \dy \right)\\
&\leq t_k D_{1,p}\bigl[m_i(1+\ep)\bigr]^{2-\frac{p'}2}
\nlp p{\omega_0}^{\frac{p'}2},  
\end{align*}
where we have used \eqref{a3} and \eqref{a4}. We conclude that 
\begin{equation}\label{a6}
|B_{k1}|\leq \frac{C}{\delta t_k}+ C_1 
\sum_{i=1}^N (|\psi(\al_i)|+\ep)\bigl[m_i(1+\ep)\bigr]^{2-\frac{p'}2}, 
\end{equation}
with $C_1 = D_{1,p}\nlp p{\omega_0}^{\frac{p'} 2} \pi^{-1}.$

\paragraf{Estimate of $B_{k2}$} We estimate directly, similarly to what was done with $B_{k1}$:
\begin{align*}
|B_{k2}|
&\leq \frac{C}{\delta t_k}+
\iint\limits_{\substack{|x-y|<\delta/3\\ x\in\HH,y\in
    E^c}}\frac{|\psi(y_1)|}{\pi t_k 
|x-y|}\wt\omega(t_k,x)\wt\omega(t_k,y)\dx\dy\\
&\leq \frac{C}{\delta t_k}+\frac{\nlp\infty\psi}{\pi t_k} \iint\limits_{\substack{|x-y|<\delta/3\\ x\in\HH,y\in
    E^c}}\frac1{|x-y|}\wt\omega(t_k,x)\wt\omega(t_k,y)\dx\dy.
\end{align*}

Lemma \ref{sjml} implies in the same way that 
\begin{align*}
|B_{k2}|
&\leq
\frac{C}{\delta t_k}+\frac{D_{1,p}}{\pi t_k} 
\nlp\infty\psi\int_{E^c}\Bigl(\int\limits_{[y_1-\frac\delta3,y_1+\frac\delta3]\times\R_+}\wt\omega(t_k,x)\dx\Bigr)^{1-\frac{p'}2}
\nlp p\omt^{\frac{p'}2} \ \wt\omega(t_k,y)\dy\\
&\leq \frac{C}{\delta t_k}+\frac{D_{1,p}}\pi\nlp\infty\psi\nlp p{\omega_0}^{\frac{p'}2}\nlp1{\omega_0} \sup_{y\in E^c} \Bigl(\int\limits_{[y_1-\frac\delta3,y_1+\frac\delta3]\times\R_+}\wt\omega(t_k,x)\dx\Bigr)^{1-\frac{p'}2}.  
\end{align*}

For $y\in E^c$, the interval $[y_1-\frac\delta3,y_1+\frac\delta3]$ is of
length less than $\delta$ and included in $\R\setminus\bigcup\limits_{i=1}^N(\al_i-\frac\delta2,\al_i+\frac\delta2)$. We deduce from \eqref{a5} that 
\begin{equation*}
\int\limits_{[y_1-\frac\delta3,y_1+\frac\delta3]\times\R_+}\wt\omega(t_k,x)\dx
<\ep.  
\end{equation*}
which implies that
\begin{equation}\label{a7}
|B_{k2}|\leq \frac{C}{\delta t_k}+ C_2 \ep^{1-\frac{p'}2},   
\end{equation}
with $C_2 =   C_1 \nlp\infty\psi \nlp1{\omega_0}$.
\medskip

Collecting the estimates for $B_{k1}$ and $B_{k2}$ (relations \eqref{a6} and \eqref{a7}) yields the following bound for $B_k$:
\begin{equation}\label{a10}
|B_k|\leq \frac{C}{\delta t_k}+ C_1 \Bigl\{ \nlp\infty\psi \nlp1{\omega_0} \ep^{1-\frac{p'}2}+ 
\sum_{i=1}^N (|\psi(\al_i)|+\ep)\bigl[m_i(1+\ep)\bigr]^{2-\frac{p'}2}\Bigr\}. 
\end{equation}
 
Take the $\limsup$ as $k \to \infty$ above to obtain:
\[\limsup_{k\to\infty}|B_k| \leq C_1 \Bigl\{ \nlp\infty\psi \nlp1{\omega_0} \ep^{1-\frac{p'}2}+ \sum_{i=1}^{\infty} (|\psi(\al_i)|+\ep)\bigl[m_i(1+\ep)\bigr]^{2-\frac{p'}2}\Bigr\}.\]
Next, send $\ep \to 0$ in order to reach the desired conclusion.
\end{proof}

\subsection{Large time asymptotics}
We will now make use of the equation for $\omt$ given in relation
\eqref{1}  together with Proposition \ref{nonlinterm} to deduce an
inequality for the limit measure $\mu$, given by \eqref{muything}. 
Surprisingly, this estimate alone will be sufficient to deduce the main result of
this paper, Theorem \ref{thdirac}. Let us begin with an outline of the proof of \eqref{muything}.
One begins with the equation for the evolution for $\omt$ \eqref{1}, taking the product with a fixed test function and integrating in space. The resulting equation has three terms. 
The first one, when integrated from $0$ to $t$, is uniformly bounded in $t$. Now, if
$\dive\bigl[y\omty\bigr]$ is weakly convergent as $t\to\infty$, then the
integral in time of the second term will, in principle, diverge like $\log t$ as $t\to\infty$. As for the third term, it is not difficult to see that it
is $\mathcal{O}(1/t)$. The dominant part of the third term must balance the logarithmic blow-up in time of the second term. The aim of Proposition \ref{nonlinterm} is precisely to estimate this dominant part of the third term.     

We will begin with a lemma, relating asymptotics on the linear part of the evolution equation for $\wt\omega$  \eqref{1} to the nonlinear part. To this end fix $\psi \in C^0(\R)$ and define the quantities
\begin{equation} \label{Apsi}
A[t;\psi] \equiv \int_\HH \psi(y_1)y_1\wt\omega(t,y)\dy, \mbox{ and }
\end{equation}
\begin{equation} \label{Bpsi}
B[t;\psi] \equiv \int_\HH \psi(y_1)\frac{\wt u_1 (t,y)}{t} \wt\omega(t,y)\dy.
\end{equation}
Note that, as the support of $\wt \omega$ is contained in a compact set
independent of $t$, it will not matter whether the support of $\psi$ is compact.

\begin{lemma} \label{linterm}
 The following estimate holds:
\begin{equation}\label{a22}
\limsup_{t\to\infty} (B[t;\psi] - A[t;\psi]) \geq 0.
\end{equation}
\end{lemma}

\begin{proof}
Let $\phi \in C^1(\R)$ be a primitive of $\psi$ so that $\phi^{\prime} = \psi$. Define 
\begin{equation*}
  f(t) \equiv \int_\HH\phi(y_1)\omty\dy,
\end{equation*}
a bounded function, since $\omt(t,\cdot)$ is bounded in $L^1$. Differentiating $f$ with respect to $t$ and using the equation \eqref{1} for $\omt$ 
we get, after integration by parts,
\[ f'(t)=\int\phi(y_1)\partial_t\omty\dy
= \frac1t\int\phi(y_1)\dive(y\omt)\dy
-\frac1{t^2}\int\phi(y_1)\dive(\wt u\omt)\dy \]
\[= \frac1{t} \int\psi(y_1) \frac{\wt u_1}{t}\omt\dy -\frac1t \int\psi(y_1)y_1\omt\dy 
\equiv \frac1{t}B[t;\psi] -\frac1t A[t;\psi]. \]
 
Integrating from $t$ to $t^2$ we obtain:
\begin{equation*}
  f(t^2)-f(t)= \int_t^{t^2}\frac{B[s;\psi] - A[s;\psi]}s\ds. 
\end{equation*}
 
Let $L = \limsup_{s\to\infty} (B[s;\psi] - A[s;\psi])$. Then, for any $\ep > 0$, there exists $M>0$ such that, if $s>M$ then $B[s;\psi] - A[s;\psi] < L+\ep$. In particular, if $t>M$ above then
\[f(t^2)-f(t) < (L+\ep)\log t,\]
so that
\[0 = \lim_{t \to \infty} \frac{f(t^2)-f(t)}{\log t} \leq L+\ep.\]
The result follows by taking $\ep \to 0$.
\end{proof}

Let us now impose a major hypothesis on the flow, namely that there exists a {\bf unique} asymptotic velocity density, so that 
\[\wt\omega(t,\cdot) \rightharpoonup \mu\otimes\delta(x_2),\]
as $t\to \infty$. We use Lemma \ref{l1} to write 
\[\mu = \nu + \sum_{i=1}^{\infty} m_i\delta_{\alpha_i}.\]

Then, for any $\psi \in C^0(\R)$, it follows that
\[A[t;\psi] \to \scal{y_1 \mu}{\psi(y_1)},\]
as $t\to\infty$. Next use Proposition \ref{nonlinterm} to deduce that 
\begin{equation*}
  \limsup_{t \to \infty} 
|B(t;\psi]| 
\leq D \nlp p{\omega_0}^{\frac{p'} 2} \sum_{i=1}^{\infty} m_i^{2-\frac{p'}{2}} | \psi(\alpha_i) |.
\end{equation*}
We therefore deduce from Lemma \ref{linterm} that:
\[\scal{y_1 \mu}{\psi(y_1)} \leq   
D \nlp p{\omega_0}^{\frac{p'} 2} \sum_{i=1}^{\infty} m_i^{2-\frac{p'}{2}}| \psi(\alpha_i) |.\]
Exchanging $\psi$ for $-\psi$ yields:
\begin{equation} \label{muything}
\bigl|\scal{y_1 \mu}{\psi(y_1)}\bigr|\leq   
D \nlp p{\omega_0}^{\frac{p'} 2} \sum_{i=1}^{\infty} m_i^{2-\frac{p'}{2}}| \psi(\alpha_i) |.
\end{equation}
The relevant fact is that the exponent $2 - \frac{p'}{2} > 1$.

Let $\delta_P$ denote the Dirac delta measure at position $P$. We are now ready to re-state our main result, giving a more precise formulation of Theorem \ref{mainthm1}.

\begin{theorem}\label{thdirac}
Suppose that the nonnegative initial vorticity $\omega_0 \in L^p_c(\HH)$ is such that there exists a {\bf unique} asymptotic velocity density $\mu$ associated to $\omega_0$. Then $\mu$ must be of the form:
\begin{equation*}
\mu=\sum_{i=1}^\infty m_i\ \delta_{\al_i} \end{equation*}
where:
\begin{enumerate}
\item \label{tha} $\al_i\neq\al_j$ if $i\neq j$ and $\al_i\to0$ as
  $i\to\infty$; 
\item \label{thb} the masses $m_i$ are nonnegative and verify $\sum_{i=1}^\infty m_i = \nlp1{\omega_0}$;
\item \label{thd} for all $i$, $\al_i\in[0,M]$,  where $M=\|u\|_{L^\infty([0,\infty)\times\HH)}$;
\item \label{the} there exists a constant $D>0$, depending solely on $p$, such that, for all $i$ with $m_i\neq0$ we have
\[\al_i\leq D\nlp p{\omega_0}^{\frac{p'}2}\ m_i^{1-\frac{p'}2}.\] 
\end{enumerate}

Furthermore, there exists $i_0$ such that $\alpha_{i_0}\neq 0$ and $m_{i_0}\neq0$.
\end{theorem}

We will need two lemmas before we give the proof of Theorem \ref{thdirac}.

\begin{lemma}\label{l2}
Let $m_i\delta_{\alpha_i}$ be a Dirac from the discrete part of
$\mu$. The following inequality holds true:
\begin{equation*}
\al_i\leq D\nlp p{\omega_0}^{\frac{p'}2}m_i^{1-\frac{p'}2}.  
\end{equation*}
\end{lemma}
\begin{proof} 
Eventually changing the order in the summation of the Diracs, we can assume that
$i=1$. Furthermore, the conclusion is trivial if $\al_1=0$, so we can
assume that $\al_1>0$ as well. Let $\ep>0$ be fixed. There exists
$\delta\in(0,\al_1)$ such that the following inequality holds:
\begin{equation*}
\mu([\al_1-\delta,\al_1+\delta])\leq m_1+\ep.  
\end{equation*}

If $\al_i\in [\al_1-\delta,\al_1+\delta]$, $i\geq2$, then
$m_1\delta_{\al_1}+m_i\delta_{\al_i}\leq\mu$ on
$[\al_1-\delta,\al_1+\delta]$, so we must have that $m_i\leq\ep$.

Let $\psi\in C^0(\R)$ be a nonnegative function supported in
$(\al_1-\delta,\al_1+\delta)\subset\R_+$ which attains its maximum at $\al_1$. By (\ref{muything}) and using the nonnegativity of $\mu$ and $y_1\psi(y_1)$ we find
\begin{equation*}
m_1\al_1\psi(\al_1)
\leq \scal{\mu}{y_1\psi(y_1)}
\leq D\nlp p{\omega_0}^{\frac{p'}2}\bigl[\psi(\al_1)m_1^{2-\frac{p'}2}
+ \sum_{i=2}^\infty\psi(\al_i)m_i^{2-\frac{p'}2}\bigr].  
\end{equation*}

We observed that if $\al_i\in (\al_1-\delta,\al_1+\delta)$, $i\geq2$, then
$m_i\leq\ep$. If $\al_i\not\in (\al_1-\delta,\al_1+\delta)$ then $\psi(\al_i)=0$. In both cases
\begin{equation*}
\psi(\al_i)m_i^{2-\frac{p'}2}\leq \psi(\al_i)\ep^{1-\frac{p'}2}m_i
\leq \psi(\al_1)\ep^{1-\frac{p'}2}m_i.
\end{equation*}
We infer that
\begin{equation*}
m_1\al_1\psi(\al_1) \leq D\nlp p{\omega_0}^{\frac{p'}2}\psi(\al_1)
\bigl[m_1^{2-\frac{p'}2}
+ \ep^{1-\frac{p'}2} \sum_{i=2}^\infty m_i\bigr],   
\end{equation*}
that is 
\begin{equation*}
m_1\al_1\leq D\nlp p{\omega_0}^{\frac{p'}2}
\bigl[m_1^{2-\frac{p'}2}
+ \ep^{1-\frac{p'}2} \sum_{i=2}^\infty m_i\bigr].   
\end{equation*}

Letting $\ep\to0$ we get that
\begin{equation*}
m_1\al_1\leq D\nlp p{\omega_0}^{\frac{p'}2}
m_1^{2-\frac{p'}2}
\end{equation*}
which implies the desired result.
\end{proof}

\begin{lemma}\label{l3}
Suppose that $\mu$ has no discrete part in some interval $(a,b)\subset\R\setminus\{0\}$. Then
$\mu\bigl|_{(a,b)}=0$.
\end{lemma}

\begin{proof} 
Let $\psi\in C^0(\R)$ with support in $(a,b)$. According to the hypothesis,
\begin{equation*}
\supp\psi\cap \{\al_1,\al_2,\dots\}=\emptyset 
\end{equation*}
so that, for this choice
of $\psi$, the right-hand side of \eqref{muything} vanishes. Therefore 
(\ref{muything}) implies 
\begin{equation*}
 \scal{\mu(y_1)}{y_1\psi(y_1)}=0 
\end{equation*}
that is
\begin{equation*}
  y_1\mu\bigl|_{(a,b)}=0
\end{equation*}
which implies the desired conclusion by recalling that $0\not\in (a,b)$.
\end{proof}

\begin{proof}[Proof of Theorem \ref{thdirac}.]
We begin by noting that Lemma \ref{l2} implies that
$\al_i\xrightarrow{i\to\infty}0$. Indeed, $\sum_{i=1}^\infty m_i<\infty$
implies that $m_i\xrightarrow{i\to\infty}0$. According to the conclusion
of Lemma \ref{l2} this immediately implies that
$\al_i\xrightarrow{i\to\infty}0$. 

Next, observe that Lemma \ref{l3} implies that the continuous part $\nu$
vanishes. Indeed, $\supp\nu\subset[0,\infty)$ since
$\supp\mu\subset[0,M]$. If $\al>0$, as $\al$ is not an accumulation
point of the set $\{\al_1,\al_2,\dots\}$, there exists
$\delta\in(0,\al)$ such that
$\{\al_1,\al_2,\dots\}\cap\bigl[(\al-\delta,\al+\delta)\setminus\{\al\}\bigr]=\emptyset$.
According to Lemma \ref{l3}, the measure $\mu$ vanishes in $(\al-\delta,\al)$ and
$(\al,\al+\delta)$, so the same is true for $\nu$. Since $\nu$ is
continuous we deduce that $\nu$ must vanish in
$(\al-\delta,\al+\delta)$. We proved that $\nu$ vanishes in the
neighborhood of each point of $(0,\infty)$. This implies that $\nu$
vanishes on $(0,\infty)$. Therefore, $\nu$ vanishes on $\R\setminus\{0\}$ and is
continuous. We conclude that $\nu=0$.

We have just proved that  
\begin{equation*}
\mu=\sum_{i=1}^\infty m_i\ \delta_{\al_i}\otimes\delta_0  
\end{equation*}
and also assertion \eqref{tha} of Theorem \ref{thdirac}. Assertion  
\eqref{thb} follows from the positivity of $\mu$ (as limit of positive measures)
and from the fact that the total mass of $\mu$ is
$\nlp1{\omega_0}$. Assertion \eqref{thd} is a consequence of the support
of $\mu$ being included in $[0,M]$ and \eqref{the} is
proved in Lemma \ref{l2}. Finally, as previously noted, it was shown in \cite{Iftimie999999}
that $\int x_1 \omega(t,x)\dx\geq Ct$ for some positive constant $C$. This implies that $\int x_1
\widetilde\omega(t,x)\dx\geq C$, which in turn yields $\sum_i m_i \alpha_i =\scal{\mu}{x_1}\geq C$.
This completes the proof of Theorem \ref{thdirac}. 
\end{proof}

\section{Extensions and Conclusions}

        We begin this section with some comments regarding the results obtained here.
\begin{enumerate}

\item The only instance of use of the energy estimate in this work is the observation that, for any asymptotic velocity density $\mu$ we have $\scal{\mu}{x_1} > C >0$, which appears when proving the last part of Theorem \ref{thdirac}. The constant $C$ depends on the kinetic energy of the initial data, as was derived in \cite{Iftimie999999}. It would be interesting to know whether kinetic energy partitions itself in a way that is consistent with the partitioning of vorticity, but we were not able to prove that, at least using only the hypothesis of uniqueness of the asymptotic velocity density.         

\item We only used the hypothesis of uniqueness of the asymptotic velocity density when we derived (\ref{muything}). The estimate on the behavior of the nonlinear term given in Proposition \ref{nonlinterm} always holds, which raises the possibility of it being exploited further. 

\item The hypothesis that the initial vorticity be $p$-integrable, with
  $p>2$ is used to ensure that the velocity is globally bounded. In
  principle, with vorticity in $L^p$, $p \leq 2$, we loose control over
  the loss of vorticity to infinity, and Lemma \ref{sjml} is no longer
  true. In fact, we do not even know the correct scaling to analyze in
  this case. 
\item We could have proved a result for initial vorticities without
  distinguished sign, which would have our main theorem here as a
  special case. The distinguished sign hypothesis was used mainly because it allows us to place the hypothesis of uniqueness of the asymptotic limit on $\widetilde{\omega}$, instead of on $|\widetilde{\omega}|$, where the authors feel it would be much less plausible.

\end{enumerate}

We would like to add a remark on the choice of the scaling $x=ty$. If
the scaling $x_1\equiv ty_1$ in the horizontal direction is motivated by
the fact that the first component of the center of vorticity behaves exactly
like $O(t)$, the scaling $x_2\equiv ty_2$ is not justified because the
second component of the center of vorticity is constant. Ideally we
should not make any rescaling in the vertical direction but then we would
have to assume that  $t\omega(tx_1,x_2)$ converges weakly, which we found
excessive because of the oscillations that may appear in the
vertical direction. We could also consider an intermediate scaling of
the form  $x_2\equiv f(t)y_2$ where
$f(t)\to\infty$ as $t\to\infty$. 
This last problem is in fact equivalent to the one we consider in this paper. 
If $f$ is such a function, then the weak limits of
$\omt_f(t,y)=tf(t)\omega\bigl(t,ty_1,f(t)y_2\bigr)$ are independent of
$f$. Indeed, let  $\nu_f$ be the weak limit of $\omt_f(t,y)$ as
$t\to\infty$ and choose a test function $h\in C_0^\infty(\overline\HH)$.
Then 
\begin{align*}
  \scal{\nu_f}{h}
&=\lim_{t\to\infty}\int_\HH \omt_f(t,y)h(y)\dy\\
&=\lim_{t\to\infty}\int_\HH \omega(t,x)h\bigl(\frac{x_1}t,\frac{x_2}{f(t)}\bigr)\dx\\
&=\lim_{t\to\infty}\Bigl( \int_\HH \omega(t,x)h\bigl(\frac{x_1}t,0\bigr)\dx
+O\Bigl(\frac{\nlp\infty{\partial_2 h}}{f(t)}\Bigr)\int_\HH x_2\omega(t,x)\dx\Bigl)\\
&=\lim_{t\to\infty}\int_\HH \omega(t,x)h\bigl(\frac{x_1}t,0\bigr)\dx
\end{align*}
since we know that $\int_\HH x_2\omega(t,x)\dx=cst.$ and $f(t)\to\infty$ as $t\to\infty$.
The last term does not depend
on $f$ anymore. Here, we have made the choice $f(t)=t$ only for the sake
of simplicity. This means that we study the asymptotic behavior of
solutions in horizontal direction but not in the vertical one. 

We would like to comment on a few problems that arise naturally from the work presented here. The first is to remove the hypothesis of uniqueness of the asymptotic velocity profile, perhaps with weaker conclusions. Also, we can try to extend this line of reasoning to other fluid dynamical situations with similar geometry, such as flow on an infinite flat channel, axisymmetric flow (smoke ring dynamics), and water wave problems. We may also ask the same questions with respect to full two-dimensional scattering, allowing for vortex pairs moving off to infinity in different directions. Finally, one might try to examine the issue of actually proving the uniqueness of asymptotic velocity densities in special cases, for example, for point vortex dynamics. We have actually looked at the case of three point vortices on the half-plane, so far without success.

\renewcommand{\thesection}{\Alph{section}}
\setcounter{section}{1}
\setcounter{equation}{0}
\setcounter{theorem}{0}
\setcounter{proposition}{0}
\setcounter{lemma}{0}
\setcounter{remark}{0}
\setcounter{definition}{0}
\setcounter{corollary}{0}
\setcounter{claim}{0}

\section*{\textbf{\sc{Appendix.}}
Separation of two vortices above a flat wall}

Steady vortex pairs provide smooth examples of vorticities for which
the corresponding asymptotic velocity densities consist of a single Dirac mass. We would like to give such an example with at least two different Dirac masses in the asymptotic
velocity density. As we pointed out in the introduction, the existence of multibump solutions in this situation is an interesting open problem, but we can offer a discrete example in order to illustrate this issue. In this section we will give a sufficient condition for linear separation of two vortices above a flat wall which will in turn give us an example of unique asymptotic velocity density concentrating at two distinct Dirac masses.

Let $z_1=(x_1,y_1)$ and $z_2=(x_2,y_2)$ be two vortices above the wall
$\{y=0\}$ of positive masses $m_1$, resp. $m_2$. For notational convenience we
will assume that we start at time $t=1$ instead of $t=0$. Let $L$ be
defined by 
\begin{equation}
  \label{app1}
  L=m_1 y_1+m_2 y_2,
\end{equation}
a quantity which is conserved by the motion of the vortices. We will prove the following proposition.
\begin{proposition}\label{papp}
Suppose there exists a positive constant $M$ such that the following relations hold true:  
\begin{gather}
x_2(1)-x_1(1)>M,\label{sep1}\\
L>m_2 y_2(1)+\frac{L^2}{\pi M^3}\label{sep2}\\
\intertext{and}
\frac{m_2}{2\bigl(y_2(1)+\frac{L^2}{\pi m_2
    M^3}\bigr)}- \frac{m_1^2}{2\bigl(L-m_2y_2(1)-\frac{L^2}{\pi M^3}\bigr)}
-\frac{2\max(m_1,m_2)}{M}> 2\pi M .  \label{sep3}
\end{gather}
Then, the two vortices $z_1$ and $z_2$ linearly separate. More precisely,
\begin{equation}
  \label{2}
  x_2(t)-x_1(t)>Mt
\end{equation}
for all times $t\ge1$.
\end{proposition}
\begin{remark}
Let $m_1$, $m_2$ and $L$ be some fixed arbitrary positive constants. Then we can always find $x_1(1)$, $y_1(1)$,  $x_2(1)$, $y_2(1)$ and $M$ such that relations \eqref{app1}, \eqref{sep1}, \eqref{sep2} and \eqref{sep3} are satisfied. Indeed, we first choose $x_1(1)$ and  $x_2(1)$ such that \eqref{sep1} holds. We next note that \eqref{sep2} and \eqref{sep3} are satisfied for large
enough $M$ and small enough $y_2(1)$. For example, if $y_2(1)=0$, then
\eqref{sep3} has a left-hand side of order $M^3$ so it is verified for
$M$ large enough; and since, for that choice of $M$, it is satisfied for
$y_2(1)=0$, it will be satisfied for small enough $y_2(1)$, too. Once $y_2(1)$ and $M$ are chosen, it remains to choose $y_1(1)$ such that \eqref{app1} is satisfied for $t=1$.
\end{remark}
\begin{proof}[Proof of Proposition \ref{papp}]
It is sufficient to prove that, as long as
\eqref{2} holds, then
\begin{equation}
  \label{3}
  (x_2-x_1)'(t)\geq M.
\end{equation}
Indeed, the result then follows by a contradiction argument: if $T$ is the
first time when $x_2(T)-x_1(T)=MT$, then necessarily $T>1$ and 
\begin{equation*}
 MT=(x_2-x_1)(T)=x_2(1)-x_1(1)+\int_1^T (x_2-x_1)'>M+M(T-1)=MT
\end{equation*}
which is a contradiction.

\bigskip

We will therefore assume in the following that \eqref{2} holds and try to
prove \eqref{3}.

\medskip

It follows from the method of images that the motion of these vortices can be computed from the full plane flow due to these two vortices together with their images: 
\begin{equation*}
z_3=\wb z_1=(x_1,-y_1)\qquad\text{and}\qquad  z_4=\wb z_2=(x_2,-y_2)
\end{equation*}
with masses $m_3=-m_1$, resp. $m_4=-m_2$. Therefore, the
equations of motion are given by:
\begin{equation*}
2\pi z_1'=\frac{(z_1-z_2)^\perp}{|z_1-z_2|^2}m_2
+\frac{(z_1-z_3)^\perp}{|z_1-z_3|^2}m_3
+  \frac{(z_1-z_4)^\perp}{|z_1-z_4|^2}m_4,
\end{equation*}
i.e., 
\begin{equation}
  \label{z1p}
\begin{split}
2\pi z_1'&=2\pi(x_1',y_1')\\
&=\bigl(\frac{m_1}{2y_1},0\bigr)
+\frac{m_2}{|z_1-z_2|^2}(y_2-y_1,x_1-x_2)
+\frac{m_2}{|z_1-\wb z_2|^2}(y_1+y_2,x_2-x_1).  
\end{split}
\end{equation}
Interchanging the indexes 1 and 2 we also get
\begin{equation}
  \label{z2p}
\begin{split}
2\pi z_2'&=2\pi(x_2',y_2')\\
&=\bigl(\frac{m_2}{2y_2},0\bigr)
+\frac{m_1}{|z_1-z_2|^2}(y_1-y_2,x_2-x_1)
+\frac{m_1}{|z_1-\wb z_2|^2}(y_1+y_2,x_1-x_2).  
\end{split}
\end{equation}

Let us now estimate $y_2$. From relation \eqref{z2p} it follows that 
\begin{equation*}
2\pi y_2'=m_1(x_2-x_1)\Bigl(\frac{1}{|z_1-z_2|^2}-\frac{1}{|z_1-\wb
  z_2|^2}\Bigr) 
=\frac{m_1(x_2-x_1)4y_1y_2}{|z_1-z_2|^2 |z_1-\wb z_2|^2}.  
\end{equation*}
In view of \eqref{app1}, we can bound $m_1 y_1\leq L$ and $y_2\leq L/m_2$
so that, using also relation \eqref{2},
\begin{equation}\label{ry2}
|y_2'|\leq \frac{2L^2}{\pi m_2 |x_1-x_2|^3} 
\leq \frac{2L^2}{\pi m_2 M^3 t^3}. 
\end{equation}
We deduce that
\begin{equation*}
|y_2(t)-y_2(1)|=\Bigl|\int_1^t y_2'\Bigr|
\leq \frac{L^2}{\pi m_2 M^3}\int_1^t\frac{2}{s^3}\ds
= \frac{L^2}{\pi m_2 M^3} \bigl(1-\frac{1}{t^2}\bigr)
\leq \frac{L^2}{\pi m_2 M^3},
\end{equation*}
which implies that
\begin{equation}
  \label{y2}
  y_2(t)\leq y_2(1)+\frac{L^2}{\pi m_2 M^3}. 
\end{equation}

Next, from \eqref{z1p}, \eqref{z2p} and \eqref{app1} we have that
\begin{align*}
(x_2-x_1)'
&=\frac{1}{2\pi}\Bigl[\frac{m_2}{2y_2}- \frac{m_1}{2y_1}
+\frac{(m_1+m_2)(y_1-y_2)}{|z_1-z_2|^2}+\frac{(m_1-m_2)(y_1+y_2)}{|z_1-\wb
  z_2|^2}\Bigr]\\
&\geq \frac{1}{2\pi}\Bigl[\frac{m_2}{2y_2}- \frac{m_1^2}{2(L-m_2y_2)}
-\frac{(m_1+m_2)}{|z_1-z_2|}-\frac{|m_1-m_2|}{|z_1-\wb z_2|}\Bigr]. 
\end{align*}
Both $|z_1-z_2|$ and $|z_1-\wb z_2|$ are bounded from below 
by $|x_1-x_2|>Mt\geq
M$. Furthermore, the first two terms of the right-hand side of the last relation are decreasing
with respect to $y_2$. We therefore deduce from \eqref{y2} that
\begin{align*}
(x_2-x_1)'
&\geq \frac{1}{2\pi}\Bigl[\frac{m_2}{2\bigl(y_2(1)+\frac{L^2}{\pi m_2
    M^3}\bigr)}- \frac{m_1^2}{2\bigl(L-m_2y_2(1)-\frac{L^2}{\pi M^3}\bigr)}
-\frac{(m_1+m_2)}{M}-\frac{|m_1-m_2|}{M}\Bigr]\\
&=\frac{1}{2\pi}\Bigl[\frac{m_2}{2\bigl(y_2(1)+\frac{L^2}{\pi m_2
    M^3}\bigr)}- \frac{m_1^2}{2\bigl(L-m_2y_2(1)-\frac{L^2}{\pi M^3}\bigr)}
-\frac{2\max(m_1,m_2)}{M}\Bigr]\\
& \geq M,   
\end{align*}
where we have used \eqref{sep3}. This completes the proof.
\end{proof}

\begin{remark}
The conclusion that $x_2(t)-x_1(t)\geq Mt$, for some $M>0$, always implies the
existence of a unique asymptotic velocity density which concentrates on
a pair of Dirac masses. In order to see this, first note that, from \eqref{ry2}, 
we have that $|y_2'|= \mathcal{O}(1/t^3)$, which implies that $y_2(t)$ converges as
$t\to\infty$ and similarly for $y_1$. From the conservation of energy we have that
\begin{equation*}
  2m_1m_2\log\frac{|z_1-z_2|}{|z_1-\wb z_2|}-m_1^2\log (2y_1)-m_2^2\log (2y_2)
\end{equation*}
is constant in time. Since $x_2(t)-x_1(t)\geq Mt$ we also know that
$\frac{|z_1-z_2|}{|z_1-\wb z_2|}\to1$ as
$t\to\infty$. We deduce that $\lim\limits_{t\to\infty}y_2(t)\neq0$ and
$\lim\limits_{t\to\infty}y_1(t)\neq0$. Now, from relations \eqref{z1p}
and \eqref{z2p} we immediately obtain that both $x'_1$ and $x_2'$
converge to a finite limit given by
\begin{equation*}
\alpha_1 \equiv \lim _{t\to\infty}x'_1(t)= \frac{m_1}{4\pi\lim
\limits  _{t\to\infty}y_1(t)} \quad\text{and}\quad 
\alpha_2 \equiv \lim _{t\to\infty}x'_2(t)= \frac{m_2}{4\pi\lim
\limits  _{t\to\infty}y_2(t)}.
\end{equation*}
Observe next that $\lim\limits_{t\to\infty}\frac{x_1(t)}{t}=\lim\limits
_{t\to\infty}x'_1(t)=\alpha_1$ and similarly for
$\frac{x_2(t)}{t}$. Finally, let us remark that the rescaled vorticity
is given in this case by $m_1\delta_{z_1/t}+m_2\delta_{z_2/t}$ so that
it clearly converges weakly to
$\bigl(m_1\delta_{\alpha_1}+m_2\delta_{\alpha_2}\bigr)\otimes\delta_0$.
Moreover, $x_2(t)-x_1(t)\geq Mt$ implies that $\alpha_2-\alpha_1\geq M >0$.
\end{remark}

\vskip\baselineskip
{\footnotesize  {\it Acknowledgments:} The first author wishes to thank Tom Sideris for many helpful discussions. The authors also wish to thank the generous hospitality of the Depts. of Math. of U.C. Santa Barbara  and UNICAMP (D.I.) and of Univ. Rennes I (H.J.N.L and M.C.L.F.) where parts of this work were accomplished. This research has been supported in part by the UNICAMP Differential Equations PRONEX and FAPESP grants \# 00/02097-1 and \# 02/02370-5.}

\providecommand{\bysame}{\leavevmode\hbox to3em{\hrulefill}\thinspace}  

\vskip\baselineskip
\noindent
{\sc
Drago\c s Iftimie\\
IRMAR, Univ. Rennes 1, Campus de Beaulieu, 35042 Rennes, France\\
and\\
Centre de Math\'ematiques, Ecole Polytechnique, 91128 Palaiseau, France
\\}
{\it E-mail address:} iftimie@math.polytechnique.fr

\vspace{.1in}
\noindent
{\sc
Milton C. Lopes Filho\\
Departamento de Matematica, IMECC-UNICAMP.\\
Caixa Postal 6065, Campinas, SP 13083-970, Brasil
\\}
{\it E-mail address:} mlopes@ime.unicamp.br

\vspace{.1in}
\noindent
{\sc 
Helena J. Nussenzveig Lopes\\
Departamento de Matematica, IMECC-UNICAMP.\\
Caixa Postal 6065, Campinas, SP 13083-970, Brasil
\\}
{\it E-mail address:} hlopes@ime.unicamp.br

\end{document}